\documentclass[12pt,preprint]{aastex}

\begin{document}
\title{Collapsars in Three Dimensions}

\author{Gabriel Rockefeller\altaffilmark{1,2}, Christopher L.
  Fryer\altaffilmark{1,3} and Hui Li\altaffilmark{2}}

\altaffiltext{1}{Department of Physics, The University of Arizona,
  Tucson, AZ 85721}
\altaffiltext{2}{Theoretical Division, LANL, Los Alamos, NM 87545}
\altaffiltext{3}{Computer \& Computational Sciences Division, LANL,
  Los Alamos, NM 87545}

\begin{abstract}
  We present the results of 3-dimensional simulations of the direct
  collapse to a black hole of a rotating, 60\,M$_\odot$, zero
  metallicity, population III star.  Because the structure of this
  star (angular momentum, density and temperature profiles) is similar
  to many collapsar gamma-ray burst progenitors, these calculations
  have implications beyond the fate of population III stars.  These
  simulations provide a first 3-dimensional look at a realistic
  collapsar progenitor, and the results are very different from any
  previous 2-dimensional calculations.  If the angular momentum of the
  progenitor is high, non-axisymmetric instabilities in the collapsing
  core cause spiral structures to form, and these structures shape
  later outflows.  These outflows are driven by the imbalance between
  viscous heating and inefficient neutrino cooling and ultimately
  develop into a $10^{52}$\,erg explosion.  Without magnetic fields,
  this collapse will not produce relativistic jets, but the explosion
  is indeed a hypernova.  We conclude with a discussion of the
  implications of such calculations on the explosions,
  nucleosynthesis, neutrino flux and gravitational wave emission from
  the collapse of massive stars.
\end{abstract}


\section{Introduction}

In the last two decades, observations have brought us a wealth of
information about gamma-ray bursts (GRBs), transforming these
phenomena from enigmatic curiosities to well-constrained astrophysical
objects.  We now know much about the immediate surroundings, host
galaxies, redshift distribution, energetics, and beaming of these
bursts \citep[e.g.][]{Sol05}.  These observations provide a number of
strong constraints on any theoretical proposal for the engine behind
these explosions.  The extreme GRB energies ($>10^{51}$\,erg) and
rapid observed variability ($\sim$\,ms) led theorists to propose
models that invoked accretion onto compact objects
\citep*[see][]{Nar92}, and very quickly a range of formation scenarios
were proposed to produce these accretion disks \citep*{Pop99,Fry99}.

One such model invoked the collapse of the core of a rotating massive
star down to a black hole \citep{Woo93}.  As the rest of the star
falls onto the core, its angular momentum causes it to hang up in a
disk, producing a rapidly-accreting ($\gtrsim 0.1 {\rm
  M_\odot\,s^{-1}}$) black hole.  It is the accretion of this stellar
material that is believed to power the GRB either through
neutrino/anti-neutrino annihilation or winding of magnetic fields.
With the discovery of supernovae-like light-curves concurrent and
cospatial with long-duration GRBs, this ``collapsar'' model became one
of the leading engines behind this class of GRBs.

Most of the work studying the collapsar model has focused on achieving
the conditions (collapse to black hole, sufficient angular momentum to
form a disk, etc.) required to produce an accreting black hole engine.
Very little work has focused on studying the collapse and explosion
process itself.  \citet{Mac99} made the first such calculations, using
a 2-dimensional, axially-symmetric variant of Prometheus
\citep*{Arn89,Fry91}.  They used simplified physics---an equation of
state for electron-degenerate matter \citep*{Blin96} with a functional
term to approximate nuclear statistical equilibrium and neutrino
cooling mimicked by a single cooling term.  They immediately collapsed
the entire iron core to a black hole and placed an inner absorbing
boundary at a radius of 50\,km to mimic that black hole.  They began
with the helium cores of several model stars and then artifically set
the angular momentum profile to ensure that the centrifugal force at
collapse would be $\leq 0.02$ times the gravitational force.

\citet{Pro03} used the same equation of state and cooling routine in
the magnetohydrodynamic code Zeus-2D to model this collapse in axial
symmetry all the way down to $1.5 R_{\rm S}$ (where $R_{\rm S} =
2GM/c^2$ is the Schwarschild radius), including the effects of
magnetic fields.  They used the same 25\,M$_\odot$ progenitor with its
ad-hoc angular momentum profile from \citet{Mac99}.  And, like
\citet{Mac99}, at the onset, they collapsed the iron core down to a
black hole.

In this paper, we present the first 3-dimensional simulations of the
collapsar, following the accretion of material onto the black hole
down to the innermost stable circular orbit.  Our progenitor is a
zero-metallicity, rotating 60\,M$_\odot$ star and the rotation profile
in our collapse is taken directly from the pre-collapse progenitor.
We discover a number of features that the previous axially-symmetric
models of collapsars did not, and could not, produce.  Most notably,
we find that the initial accretion torus develops spiral wave
instabilities, leading to enhanced angular momentum transport and
viscous heating and the strong production of gravitational waves.  In
\S 2, we discuss our initial conditions and computational technique
and the general features evident in our results.  We describe the
transport of angular momentum through our simulated collapsar in \S 3
and discuss the outflows from the central accreting torus in \S 4.  We
conclude in \S 5 with a discussion of the implications of these
3-dimensional accretion instabilities, including analysis of the
gravitational waves produced in these models.

\section{Code and Generic Results}

\begin{figure}
\plotone{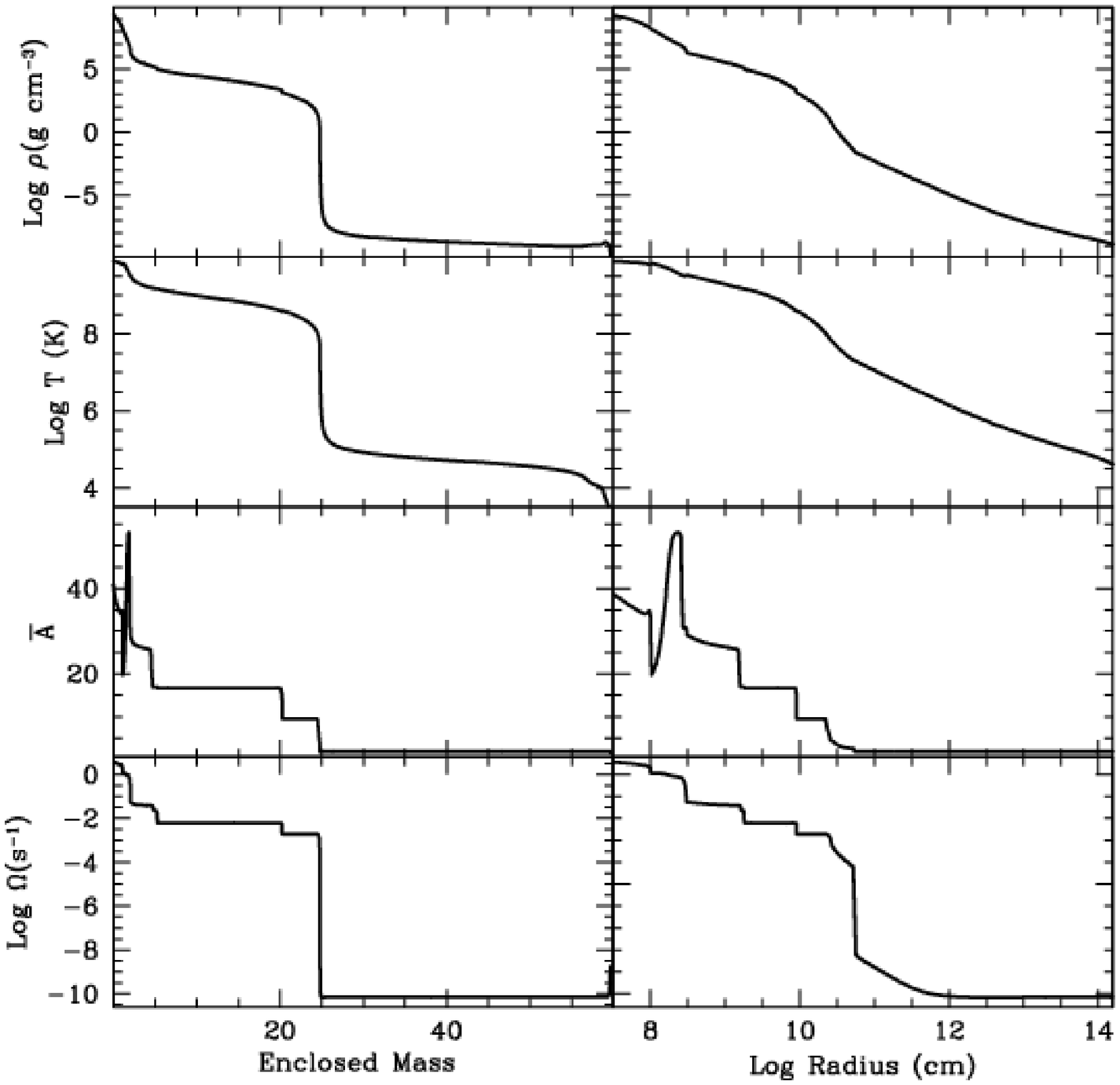}
\caption{Density, temperature, average atomic weight ($\bar{A}$), and
  angular velocity ($\Omega$) versus enclosed mass and versus radius
  for our 1-dimensional progenitor.  The peak density occurs in the
  inner few solar masses, and this material will collapse first,
  presumably forming a black hole.  The sharp jumps in the angular
  velocity (and less noticeably, but more importantly, in the density)
  at element boundaries in this star are caused by the simplistic
  mixing algorithm in the KEPLER code.  Such jumps affect the collapse
  and are an indication of the uncertainties in the initial progenitor
  that plague all stellar collapse simulations.}
\label{fig:prog1}
\end{figure}

Our model progenitor is a rotating 60\,M$_\odot$ star evolved to
collapse using the stellar evolution code KEPLER \citep{Wea78}.
Figure~\ref{fig:prog1} shows the density, temperature, average atomic
weight ($\bar{A}$), and angular velocity ($\Omega$) versus enclosed
mass and versus radius for this 1-dimensional simulation at collapse.
The stellar evolution calculation did not include magnetic torques and
the rotational velocity in the core is at the high end of possible
values \citep[for more details, see][]{Heg06}.  The peak density
occurs in the inner few solar masses, and this material will collapse
first, presumably forming a black hole.

Note the sharp jumps in the angular velocity (and less noticeably, but
more importantly, in the density) at element boundaries in this star.
This is the result of the simplistic mixing algorithm in the KEPLER
code.  More advanced mixing techniques for modeling convection in
modern stellar evolution codes \citep[e.g.][]{You05} do not produce
such sharp boundaries.  The process of mapping this 1-dimensional
model into three dimensions for use with our smoothed particle
hydrodynamics (SPH) code tends to smooth out these boundaries
slightly, but they are still quite sharp in our 3-D initial conditions
and such uncertainties in our progenitor dominate the uncertainties in
our calculation.  That is, the 60\,M$_\odot$ star we are currently
modeling may look very different than the true structure of a
60\,M$_\odot$ star.  As stellar evolution codes improve, we hope this
source of error will shrink.

The 1-dimensional structure is converted into a series of shells in
our 3-dimensional SPH setup.  The spacing between the shells is set to
be roughly equal to the spacing between the particles within a shell.
We used 2.5 million particles to model the inner 5.6\,M$_\odot$ core
of the progenitor with particle masses ranging from less than
$10^{-6}$\,M$_\odot$ to above $10^{-5}$\,M$_\odot$.  The particles are
distributed in the shells randomly \citep{FW02} so there is no
preferred axis \citep*[see][for details]{Fry06}, but this does
introduce perturbations in the density profile; these perturbations
are roughly 10\% where the density gradient is large, but less than
1\% where the density is relatively flat.  These perturbations will
seed convection in the collapsed core.

\begin{figure}
\plotone{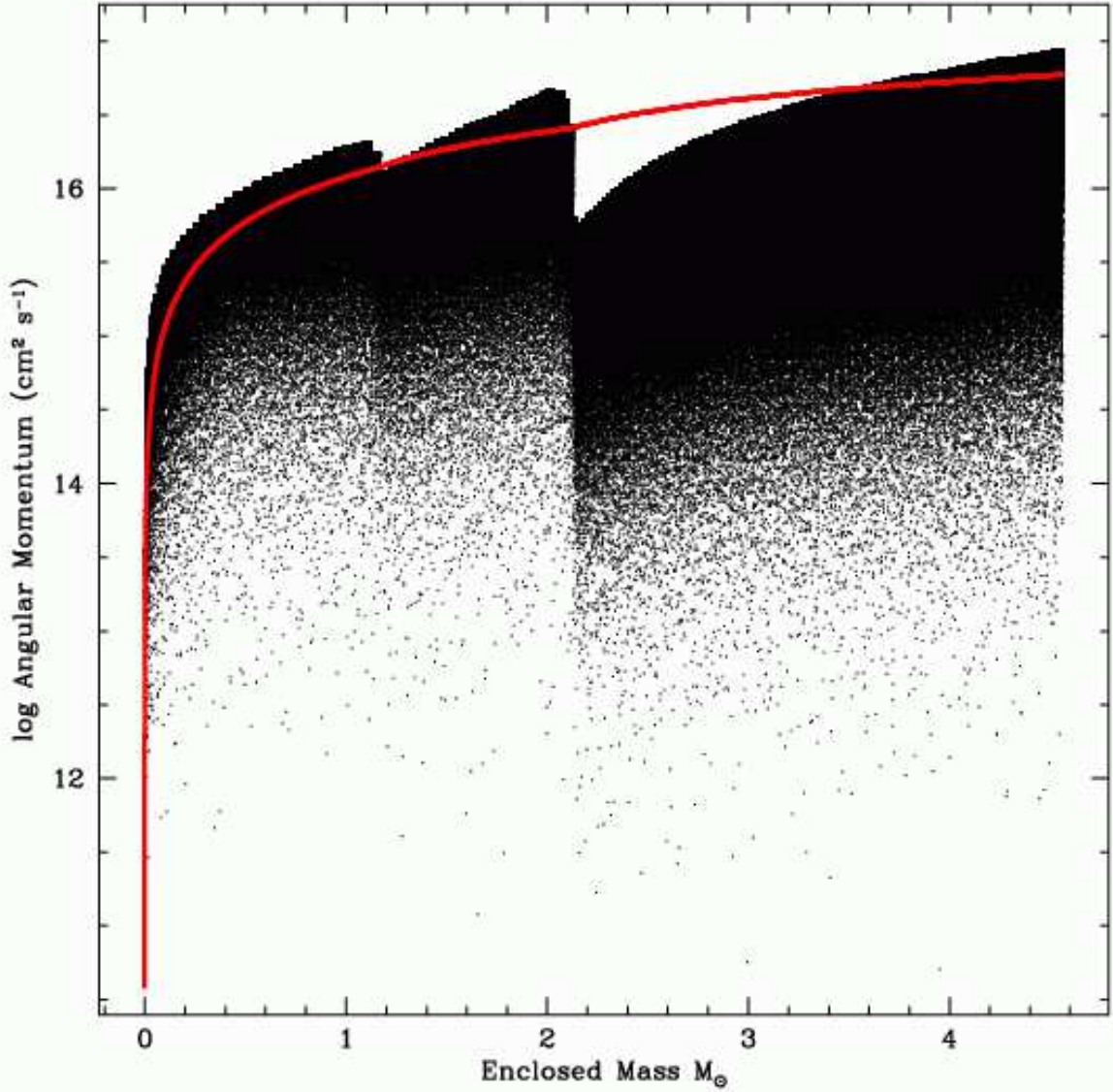}
\caption{Specific angular momentum versus enclosed mass for our 
  3-dimensional setup.  The red line shows the angular momentum needed
  for a piece of matter to be centrifugally supported at the innermost
  stable circular orbit above a black hole.  Here we have assumed the
  black hole mass and angular momentum are set to the mass and angular
  momentum of all matter interior to the point in question.  Note that
  over most of the mass, the distribution of angular momenta at a
  fixed radius implies that some particles will fall directly into the
  black hole while others will hang up in a disk.}
\label{fig:prog2}
\end{figure}

The angular velocity is added equally into each shell with an axis
along the z-axis \citep{FW02}.  Hence, the angular momentum of a given
particle in a shell is directly proportional to the square of its
distance away from the z-axis.  The specific angular momentum of each
particle versus the enclosed mass at smaller radii in the progenitor
is shown in Figure~\ref{fig:prog2}.  For guidance, we also plot the
specific angular momentum needed for a piece of matter to be
centrifugally supported at the innermost stable circular orbit above
the black hole.  Here we have assumed that the black hole mass and
angular momentum are set equal to the mass and angular momentum of all
matter interior to the point in question.  Note that over most of the
mass, the distribution of angular momenta at a given radius implies
that some particles (i.e. particles at higher latitudes above the
equatorial plane) will fall directly into the black hole while others
will hang up in a disk.

We constructed two initial models and ran two SPH simulations; the
first model uses the angular momentum profile taken directly from the
results of the KEPLER calculation (and shown in
figure~\ref{fig:prog2}), while the second model uses an angular
momentum profile where the magnitude of the angular momentum is
reduced by a factor of 10 compared to the KEPLER results.  We refer to
the first model as the ``rapidly-rotating'' model and to the second
model as the ``slowly-rotating'' model.

The collapse and resultant explosion of each star is modeled in one
continuous simulation using the smoothed particle hydrodynamics code
SNSPH \citep{Fry06}.  SNSPH uses a tree-based self-gravity module that
has been tested to accurately follow the collapse but also include the
disk formation.  It also includes a 3-flavor flux-limited diffusion
package to simulate both neutrino cooling and the absorption of
neutrinos (which affects the electron fraction of the ejecta).  We use
a coupled equation of state; at low densities we use the equation of
state by \citet{Blin96}, which was also used by \citet{Mac99} and
\citet{Pro03}, while at high densities we employ a dense equation of
state by \citet{Lat91}.  To include the effects of nuclear burning, we
include a nuclear statistical equilibrium package
\citep[see][]{Her94,Fry06}.

We do not model the initial formation of a neutron star at the center
of the collapsing progenitor; instead, we assume that the core quickly
collapses to form a black hole.  To mimic this, we begin the
simulation with an inner absorbing boundary.  The inner boundary of
the calculation is initially set at a radius of 10\,km and is allowed
to expand once the accreted mass and angular momentum are sufficient
to move the innermost stable circular orbit around the central black
hole out to larger radii.  In nature, there will be a period of time
before the core has collapsed to a neutron star, and this will change
the initial evolution of the collapse and possibly even the explosion.

\section{Angular Momentum Transport}

\begin{figure}
\plotone{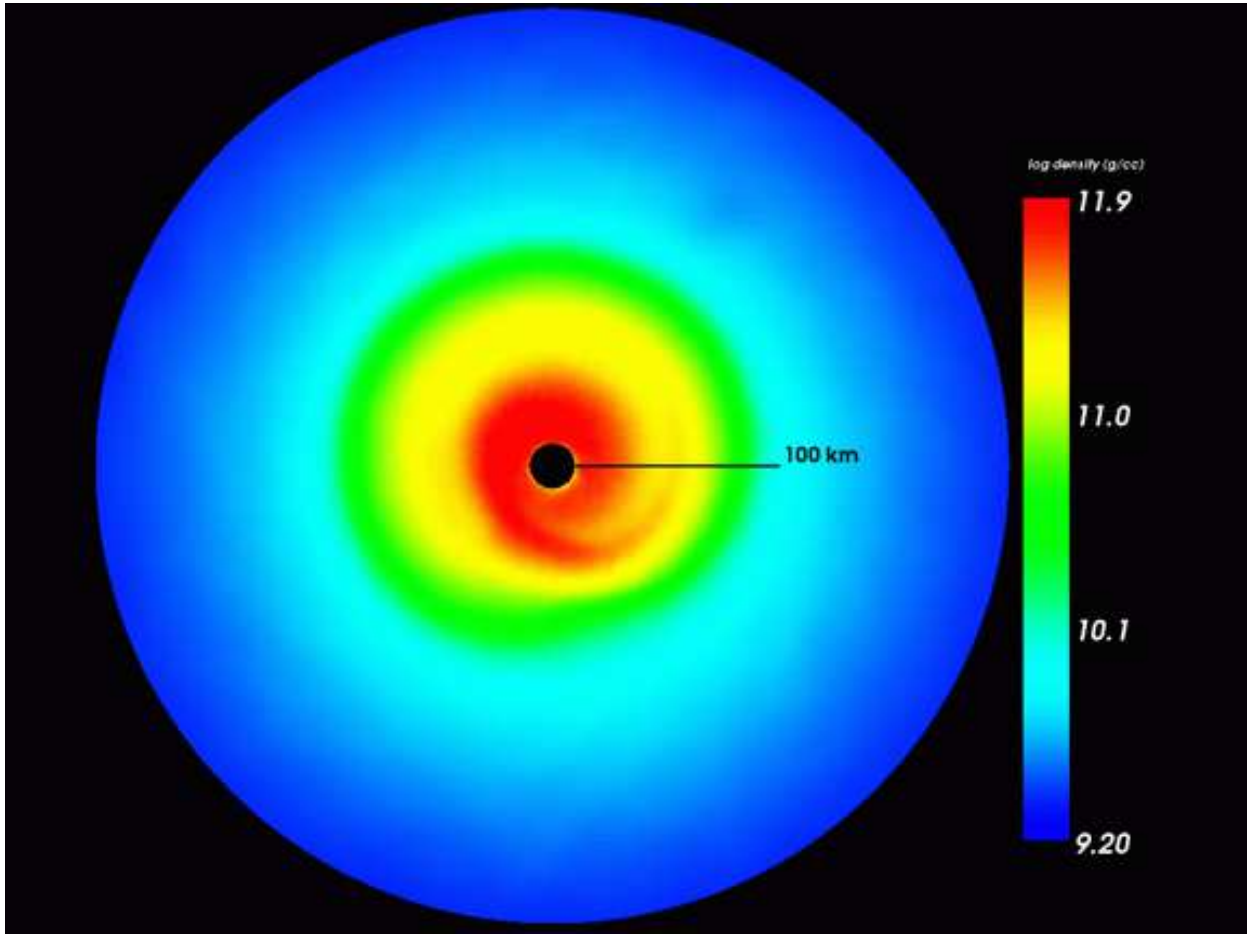}
\caption{The matter density in the equatorial plane of the rapidly-rotating 
  collapsar simulation, $0.30$ seconds after collapse.  The spiral
  wave becomes visible near the center of the collapsar $\sim 0.29$~s
  after collapse.}
\label{fig:rho1}
\end{figure}
\begin{figure}
\plotone{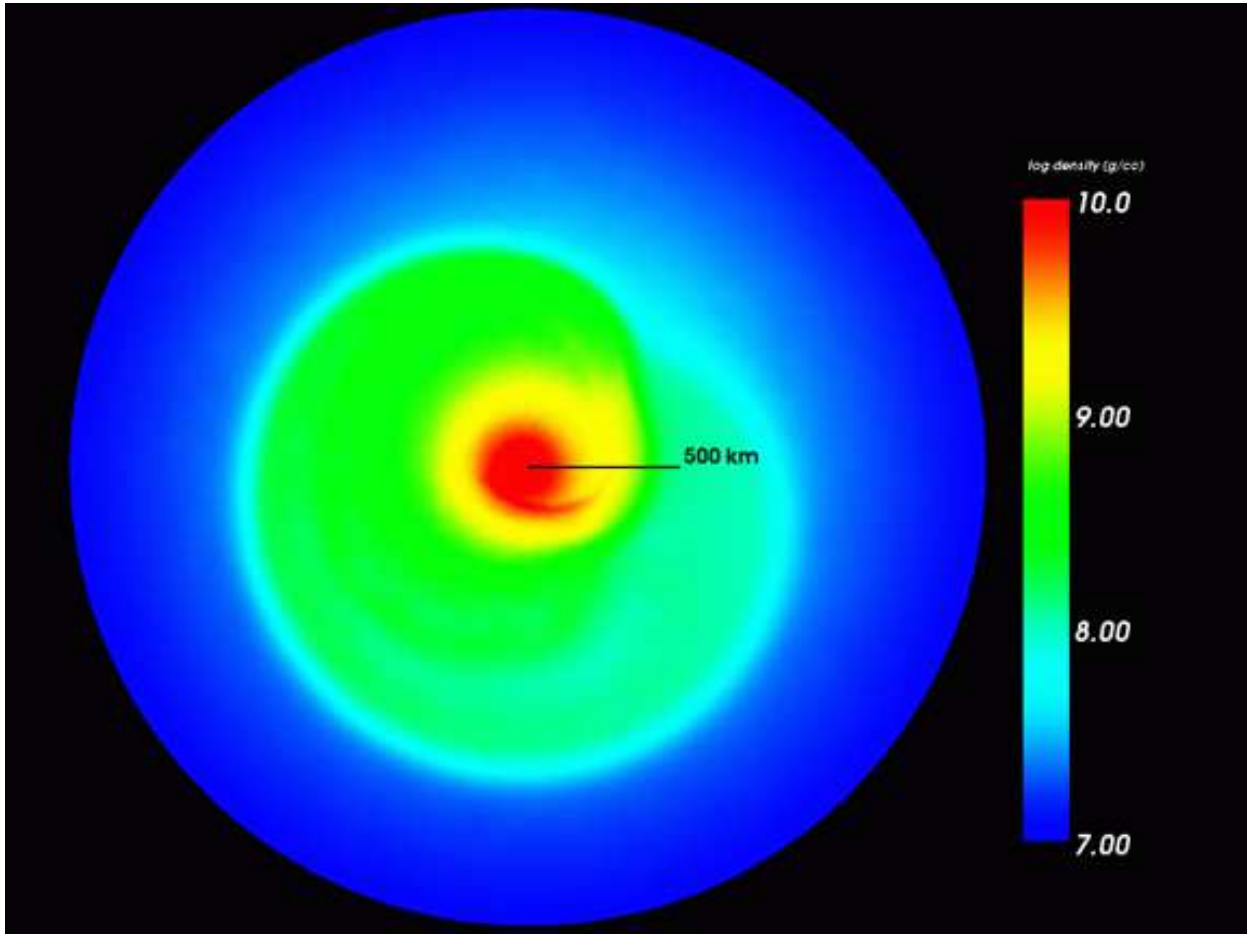}
\caption{The matter density in the equatorial plane of the rapidly-rotating 
  collapsar simulation, $0.44$ seconds after collapse.  The spiral
  wave forms near the center of the collapsar $0.29$~s after collapse
  and moves outward through the star.}
\label{fig:rho2}
\end{figure}

The material flowing toward the inner boundary carries enough angular
momentum to form a small torus at the center of the rapidly-rotating
star.  This torus is hot ($> 10^{10}$~K), and gas pressure gradients
inside it produce forces on par with the centrifugal force.  The torus
is susceptible to the formation of spiral waves
\citep*{Pap84,Gold86,Li01}; a spiral structure becomes apparent at a
time $t \simeq 0.29$~s after collapse.  Figure~\ref{fig:rho1} shows
the density in the equatorial plane of the rapidly-rotating star
0.30~s after collapse, while figure~\ref{fig:rho2} shows the density
0.44~s after collapse.

The pressure gradients associated with the spiral wave are capable of
transporting additional angular momentum beyond that transported by
fluid viscosity.  Azimuthal pressure gradients generate positive
correlations between the radial and azimuthal velocities of fluid
elements.  Compared to a ``background'' fluid element not near a
spiral wave, a fluid element ahead of the spiral wave will tend to
move with a faster azimuthal velocity and a higher outward radial
velocity, and a fluid element behind the spiral wave will tend to move
with a slower azimuthal velocity and a smaller outward (or larger
inward) radial velocity; both of these fluid elements transport
additional angular momentum outward through the star.

To characterize the transport of angular momentum during the
calculation, we follow \citet{Li01} and calculate the variation in
radial and azimuthal velocity components at points throughout the
equatorial plane of the collapsar.  We then use the correlations
between these variations to estimate a local $\alpha$ coefficient at
each point in the flow.

We begin by calculating azimuthally-averaged radial and azimuthal
velocities $\langle {v}_r \rangle$ and $\langle {v}_\phi \rangle$ in
the midplane of the simulation at a given time.  In general the actual
radial and azimuthal components of the velocity at a given point
differ from these average values; the radial velocity $v_r$ differs
from the average by $\delta v_r$ ($v_r = \langle {v}_r \rangle +
\delta v_r$), and the azimuthal velocity $v_\phi$ differs by $\delta
v_\phi$ ($v_\phi = \langle {v}_\phi \rangle + \delta v_\phi$).  The
two-dimensional $\alpha$ coefficient $\alpha_{r\phi}$ at a given point
is proportional to the product of these velocity deviations,
\begin{equation}
\alpha_{r\phi} = \frac{\rho \delta v_r \delta v_\phi}{P} ,
\end{equation}
where $\rho$ is the fluid density and $P$ is the pressure.  Positive
values of $\alpha_{r\phi}$ identify regions where radial and azimuthal
deviations from the average fluid velocity are positively
correlated---in other words, where outward transport of angular
momentum is occurring.

\begin{figure}
\plotone{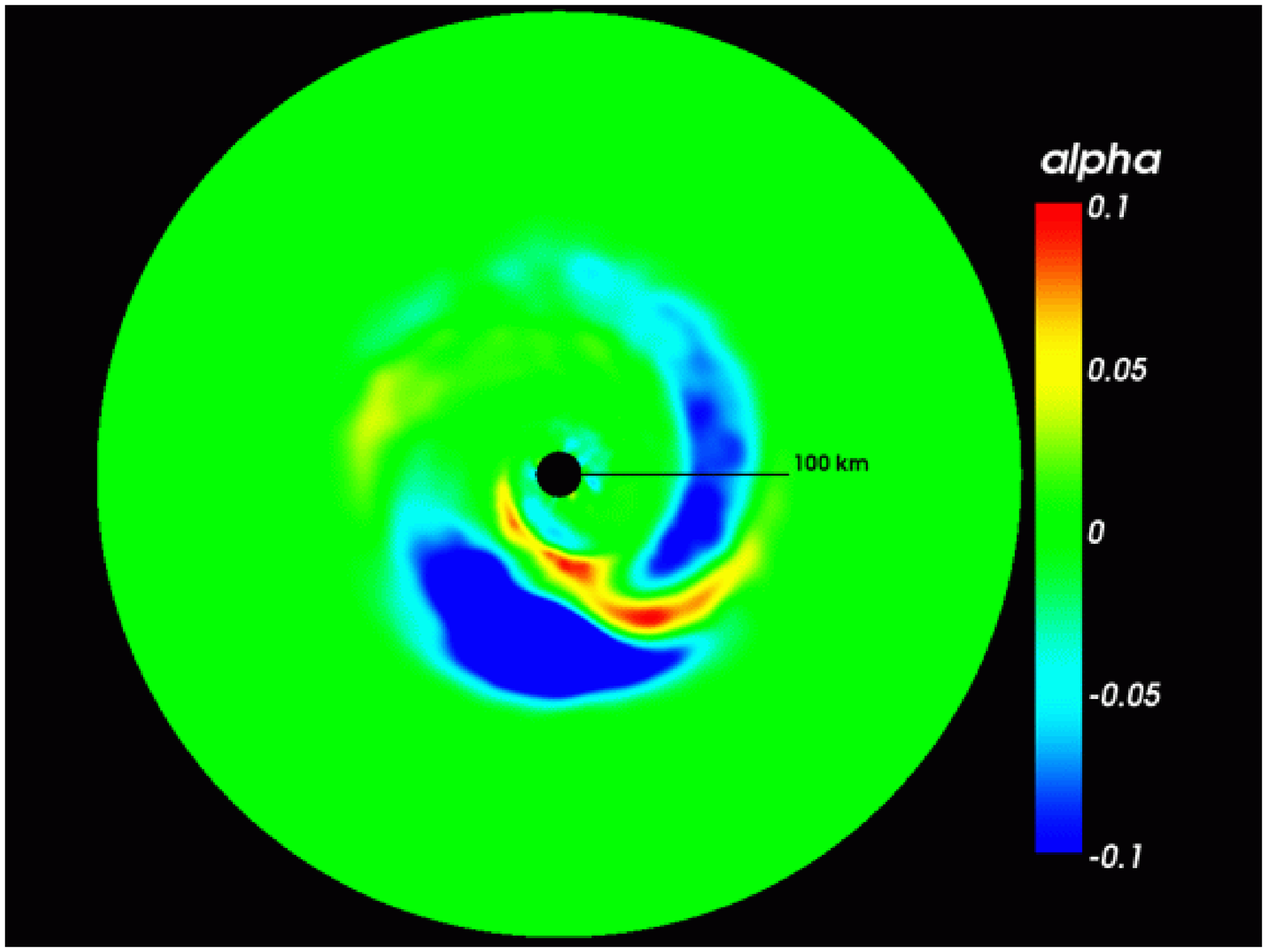}
\caption{The two-dimensional $\alpha$ coefficient $\alpha_{r\phi}$ in 
  the equatorial plane of the rapidly-rotating collapsar simulation,
  $0.30$ seconds after collapse.  The red regions (i.e. regions of
  large outward angular momentum transport) lie along the leading
  edges of the spiral wave.  The color scale was restricted to the
  range $\left| \alpha_{r\phi} \right | \leq 0.1$ to more clearly
  illustrate small spatial variation in $\alpha_{r\phi}$; the actual
  range in the data is $-0.4 \leq \alpha_{r\phi} \leq 0.1$.}
\label{fig:alpha1}
\end{figure}

\begin{figure}
\plotone{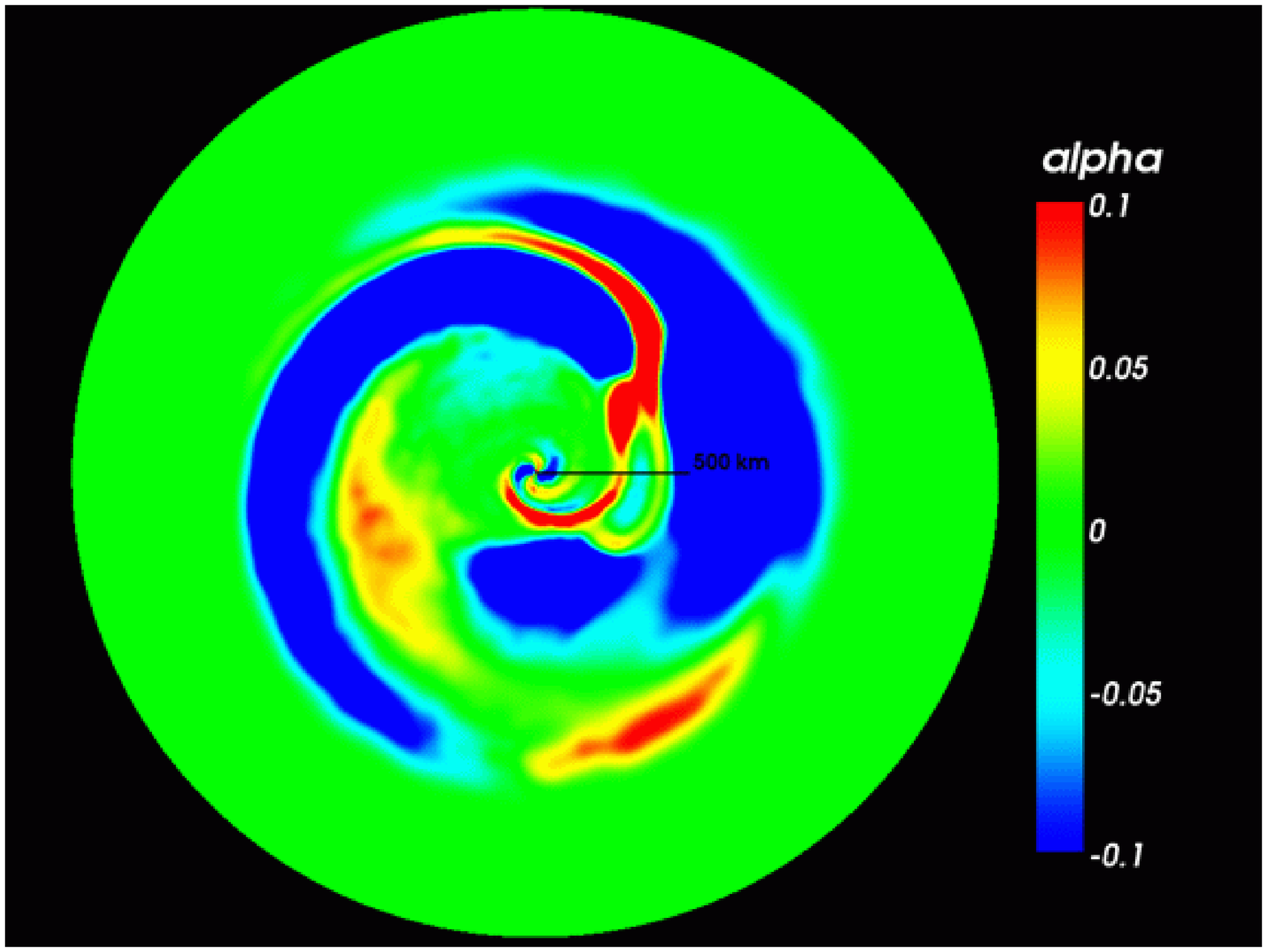}
\caption{The two-dimensional $\alpha$ coefficient $\alpha_{r\phi}$ in 
  the equatorial plane of the rapidly-rotating collapsar simulation,
  $0.44$ seconds after collapse.  The red regions (i.e. regions of
  large outward angular momentum transport) lie along the leading
  edges of the spiral wave.  The color scale was restricted to the
  range $\left| \alpha_{r\phi} \right | \leq 0.1$ to more clearly
  illustrate small spatial variation in $\alpha_{r\phi}$; the actual
  range in the data is $-0.6 \leq \alpha_{r\phi} \leq 0.4$.}
\label{fig:alpha2}
\end{figure}

Figure~\ref{fig:alpha1} shows the value of $\alpha$ in the equatorial
plane of the rapidly-rotating collapsar, 0.30~s after collapse, while
figure~\ref{fig:alpha2} shows the value of $\alpha$ 0.44~s after
collapse.  Red regions show positive correlations between azimuthal
and radial velocity fluctuations and are regions of efficient outward
transport of angular momentum.  Blue regions show negative
correlations between azimuthal and radial velocity fluctuations and
are regions of efficient inward angular momentum transport.  Green
regions, especially in regions of the star well outside the spiral
wave, indicate angular momentum transport due solely to fluid
viscosity---in our calculation, due to the artificial viscosity
implemented in our code.  The artificial viscosity in our code results
in a value of $\alpha \lesssim 0.01$ (depending on the particular
region of the flow; fluid at larger radii shows smaller velocity
fluctuations and therefore has a smaller typical $\alpha$ value),
while fluid elements near the spiral wave have values of $\alpha
\gtrsim 0.1$.  The spiral wave does produce local regions of enhanced
outward angular momentum transport, but it affects only a small
fraction of the material in the star and does not explain the
large-scale outflows we find in our simulation (though it does help
explain the asymmetric structure of those outflows).

\begin{figure}
\plotone{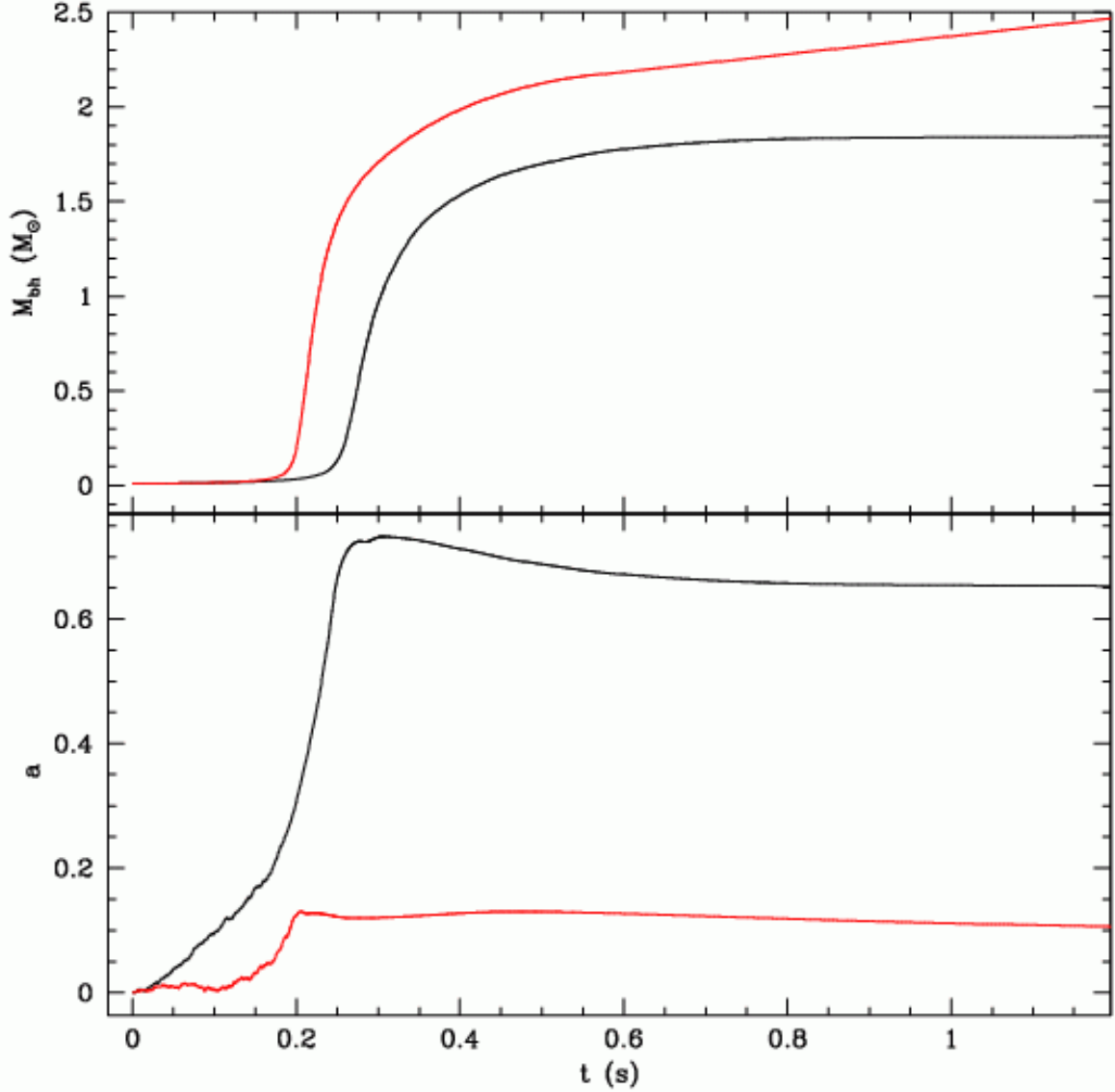}
\caption{The black hole mass $M_{\rm bh}$ and spin parameter $a$ as 
  functions of time for the rapidly-rotating (black line) and
  slowly-rotating (red line) collapsars.  A spiral wave forms in the
  center of the rapidly-rotating model, $0.29$~s after collapse;
  viscous heating near the spiral wave initiates outflows and
  eventually prevents further accretion.}
\label{fig:mdot}
\end{figure}

Figure~\ref{fig:mdot} shows the black hole mass $M_{\rm bh}$ and spin
parameter $a$ as functions of time for both the rapidly-rotating and
slowly-rotating collapsar simulations.  The accretion rate $\dot{M}$
through the inner boundary of our simulation is highest around $t =
0.28$~s after collapse, just before the spiral wave becomes clearly
visible in images of the density in the equatorial plane.  $\dot{M}$
rises rapidly in the $\sim 80$~ms leading up to the appearance of the
spiral wave and drops more gradually during the following $\sim
250$~ms.  When combined with the results from figures~\ref{fig:alpha1}
and \ref{fig:alpha2}, discussed above, this supports the conclusion
that the presence of a spiral wave in our simulated collapsar does not
significantly enhance the transport of angular momentum through the
inner region of the star.  Instead, enhanced viscous heating near the
spiral wave helps initiate large outflows that shut off accretion onto
the black hole; these outflows are described below.

\section{Outflows}

\begin{figure}

\plottwo{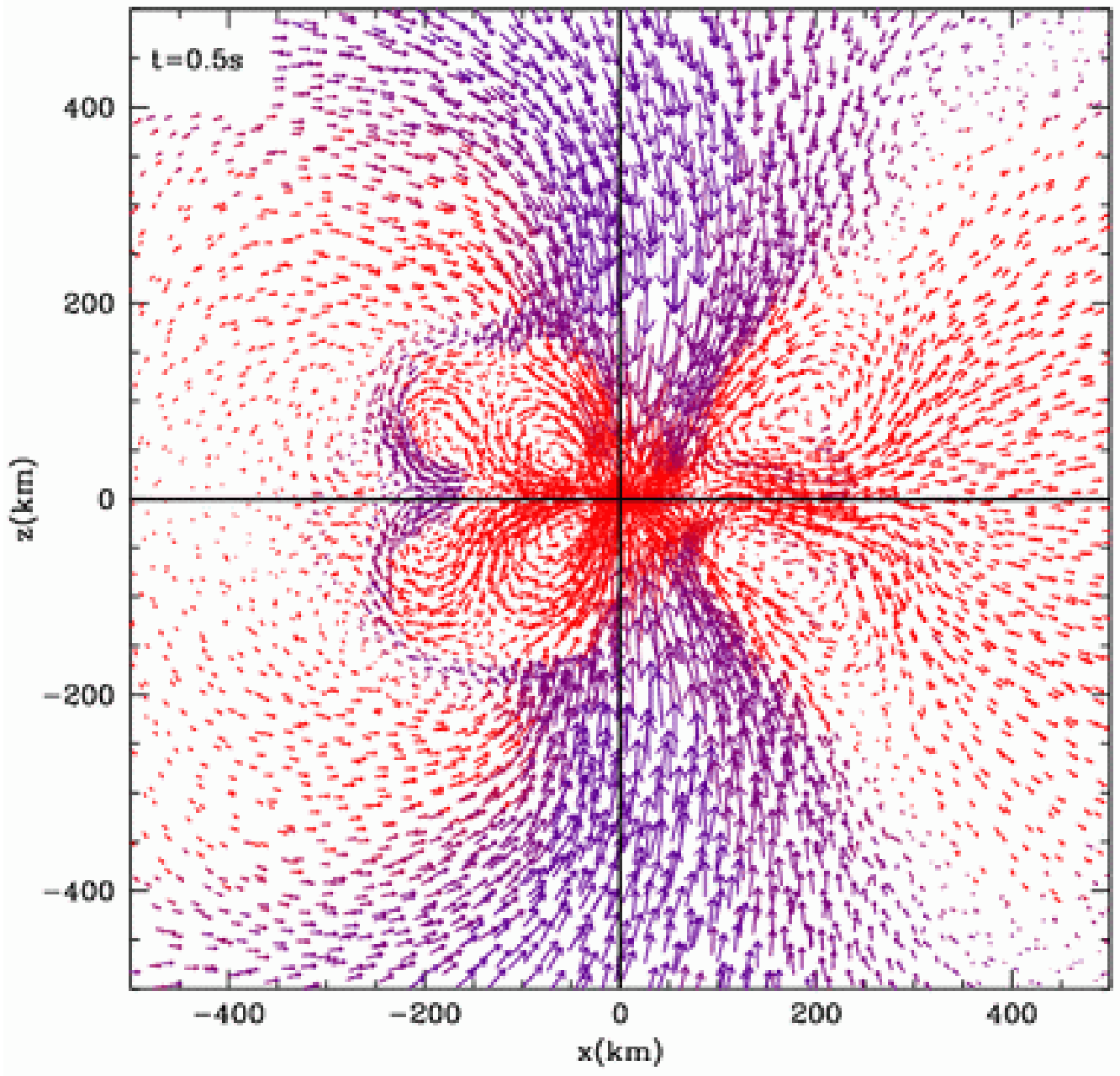}{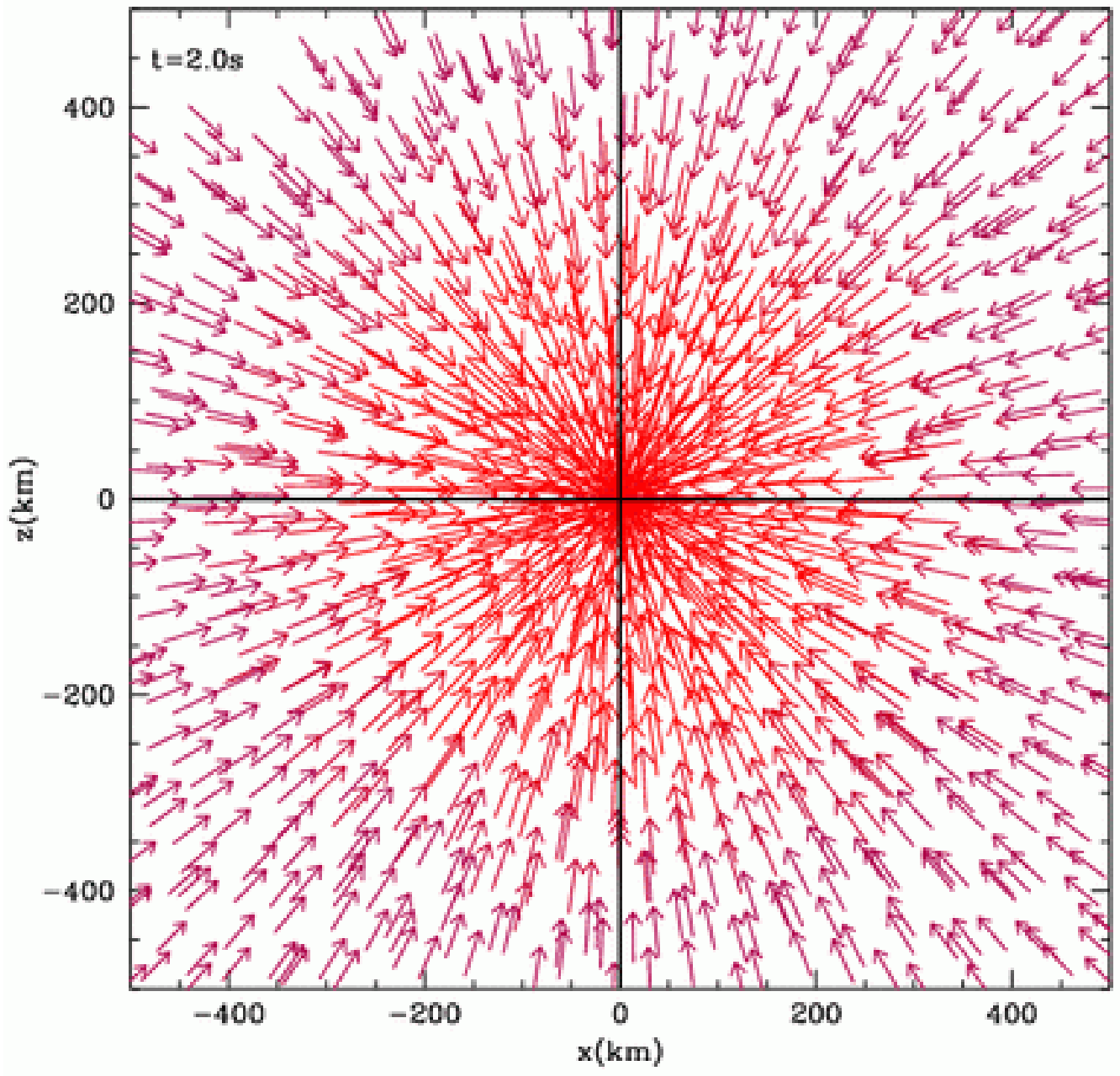}

\caption{Slices in the x-z plane of both our rapidly-rotating (left) 
  and slowly-rotating (right) models.  We show three separate
  snapshots in time for both models.  The color coding denotes the
  entropy, and the vectors denote velocity magnitude (length) and
  direction.  Although most of the material initially falls radially
  inward in our rapidly-rotating case, angular momentum causes this
  material to hang up in a torus, and heating ultimately produces
  strong outflows.  Note how different these outflows are from
  canonical disk outflows.  The convective motions in this outflow are
  more akin to a standard supernovae than an accretion disk.  The only
  difference is the fact that centrifugal support, not the hard
  surface of a neutron star, is halting the inflow of material.  By
  0.5\,s the outflow is well beyond 500\,km.  For the slowly-rotating
  model, the radial inflow persists to times later than 2.0\,s after
  collapse.}
\label{fig:out1}
\end{figure}

The intense activity in the accretion torus causes considerable viscous
heating and ultimately drives outflows.  Figure~\ref{fig:out1} shows
three snapshots in time for both the rapidly- and slowly-rotating
simulations.  The strong viscous heating in the rapidly-rotating case
leads to outflows, but because the material is accreting through all
directions (not just along the equator of rotation), the outflows
differ significantly from the canonical disk outflows.  With time,
this model eventually halts all accretion by blowing up the star.  The
low-angular-momentum star does not undergo significant heating (only
near the innermost stable circular orbit does viscous heating start to
overcome the neutrino cooling) and the material accretes directly onto
the black hole.  There is very little time variation in the accretion
profile.

\begin{figure}
\plotone{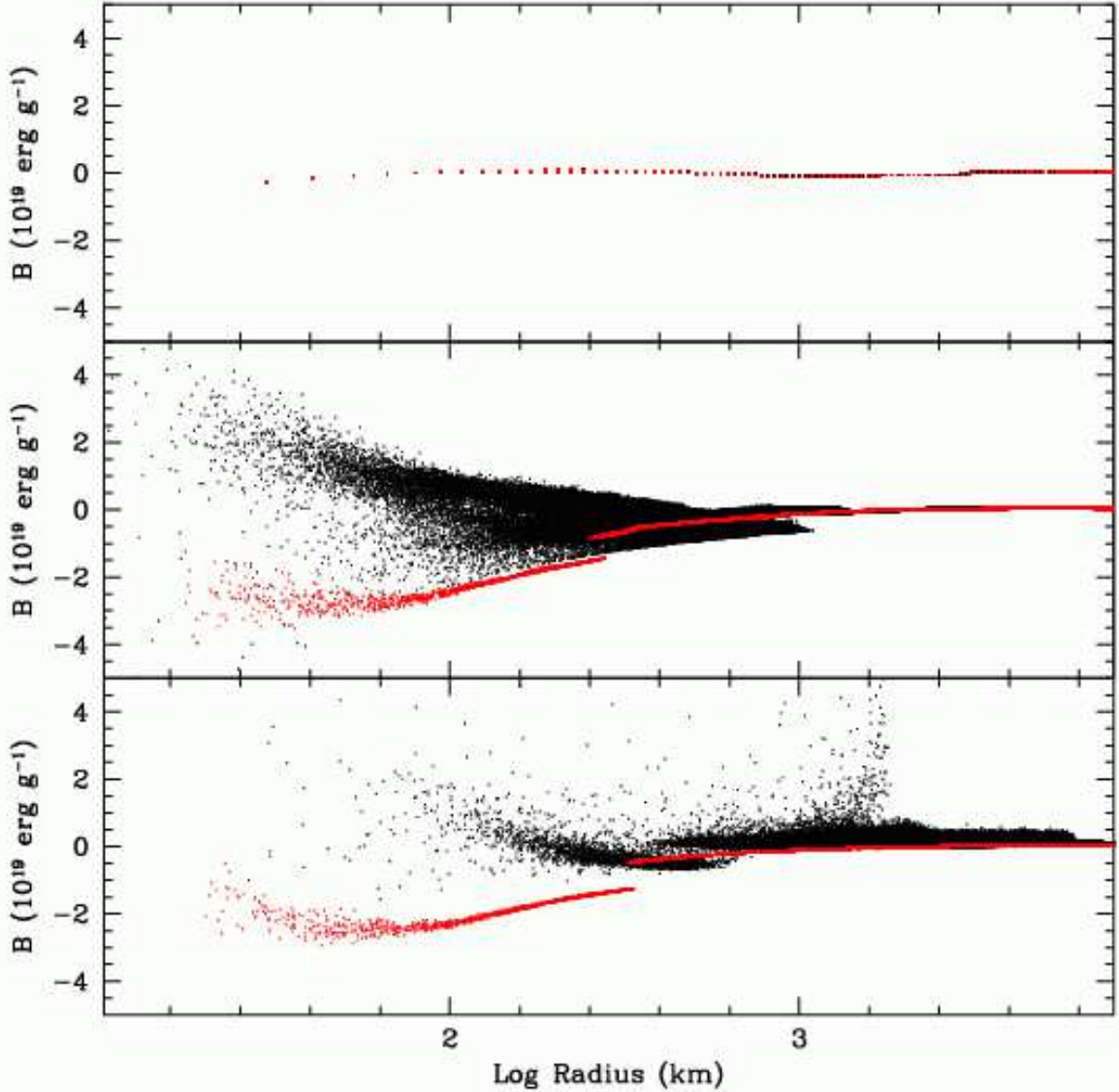}
\caption{The value of the Bernouli function (eq.~\ref{eq:bernouli}) as a
  function of radius at three different snapshots in time.  The
  rapidly-rotating (black) and slowly-rotating (red) models are
  plotted together.  Note that at time $t = 0$, the values of both
  models are nearly identical.  This is because both the enthalpy and
  gravitational potential energies for these two models are initially
  identical, and the rotational energy is not a dominant term until
  the matter collapses.  Since the rotational energy (if angular
  momentum is conserved) increases inversely with the square of the
  radius, it quickly becomes more important and viscous heating drives
  outflows.  For the rapidly-rotating model, the heating starts to
  overcome neutrino cooling as far out as 1000\,km.  For the
  slowly-rotating model, viscous heating only starts to overcome the
  neutrino cooling at a few times the innermost stable circular orbit,
  and it is never able to drive the Bernouli function positive.}
\label{fig:out2}
\end{figure}

One way to understand the outflows is to look at the Bernoulli
function $B$ for the infalling matter, 
\begin{equation}
B=H+j^2/(2 r^2) - GM_{\rm enclosed}/r,
\label{eq:bernouli}
\end{equation}
where $H$, $j$, and $r$ are, respectively, the enthalpy, specific
angular momentum and radius of a piece of matter, $G$ is the
gravitational constant, and $M_{\rm enclosed}$ is the mass interior to
that material.  \citet{Blan99} have argued that if $B$ is positive,
outflows occur.  Although this is not strictly true \citep[see][for
details]{Abram00,Blan04}, the sign of the Bernouli function remains an
ideal indicator of the existence of outflows.  Figure~\ref{fig:out2}
shows the value of this function for both our models at three
different times during the simulations.  Its value is nearly identical
for the two models for all of the matter at the beginning of the
simulation; both the enthalpy and gravitational potential energies are
identical, and the rotational kinetic energy is very low.  However,
since angular momentum is conserved, this rotational kinetic energy
grows as $r^{-2}$ as material moves inward (versus $r^{-1}$ for the
gravitational potential energy), and it quickly becomes a dominant
term.  This rotational energy is then converted to heat that then
drives the outflow.  In our rapidly-rotating progenitor, the Bernouli
number becomes positive for the bulk of disk matter out to roughly
1000\,km (over 100 Schwarzchild radii).  In comparison, the material
in the slowly-spinning progenitor does not undergo significant heating
until this matter is down at only a few times the innermost stable
orbit.  Even so, the viscous heating does not drive the Bernouli
function above zero.

Another way to understand the outflows in the rapidly-rotating model
is just through heating and cooling terms.  If the cooling in an
accretion flow is inefficient and there is some transport of energy,
it is very difficult to accrete large amounts of matter.  This is
because the matter is only marginally bound.  If it doesn't lose its
energy through neutrino emission, it only needs to gain a small amount
of energy to become unbound.  If the star is not rotating rapidly,
there will not be enough heating and the material will accrete
directly onto the black hole; conversely, if the star is rotating
sufficiently rapidly, viscous heating can drive outflows.  Neutrino
emission depends sensitively on the temperature and is not effective
until the matter achieves high temperatures, so neutrino cooling is
inefficient.  In our high-angular momentum case, there is enough
viscous heating transporting energy from the accreting material to the
infalling material to drive considerable mass ejection.

Both \citet{Mac99} and \citet{Pro03} found outflows and we expected
such outflows to occur.  But our outflows have very different
structures than those found by these two groups.  One of the reasons
for this difference is their artificial choice for the angular
momentum profile of the progenitor.  Their choice of initial angular
momentum produced a more canonical disk structure.  Our initial
conditions, taken directly from a stellar evolution model, lead to
convection that is much closer to a classic supernova model, except
that centrifugal support, not degeneracy pressure, halts the inflow.


We did not include the energy deposited through neutrino/anti-neutrino
annihilation.  However, the neutrino fluxes from our simulation are
quite low: below $10^{52}$\,erg\,s$^{-1}$ (Fig.~\ref{fig:neut}).
Using the analysis of \citet{Pop99} as a guide, we would expect less
than $10^{50}$\,erg of energy to be deposited by neutrino
annihilation.  Given that viscous heating ultimately drives a
$10^{52}$\,erg explosion in our rapidly-rotating model, neutrino
annihilation can reasonably be neglected.

We also do not include the effects of magnetic fields.  The SPH
technique does have an artificial viscosity which deposits energy
locally, mimicking the ``alpha-viscosity'' transport of angular
momentum that we might expect from a turbulent magnetic field.  But
even in our extremely altered structures near the black hole, it is
possible that some sort of dynamo will produce global magnetic fields
that drive a jet along the rotation axis like that seen by
\citet{Pro03}.  Without magnetic fields, we do not produce this jet
(and hence, do not produce a gamma-ray burst); this calculation is
still incomplete and we are far from understanding all the features of
such collapse models.

\section{Implications}

We have presented the results from two collapsar simulations of a
60\,M$_\odot$ star.  These are the first simulations of the collapsar
model in three dimensions.  The rapidly-spinning star (using the
angular momentum profile of the stellar evolution model) quickly forms
a small accretion torus which eventually ejects considerable mass in a
disk-driven wind.  This ejecta drives a hypernova-like explosion with
a total explosion energy above $10^{52}\, {\rm erg}$.  Our second
simulation, with one-tenth the total angular momentum, accreted
directly onto the nascent black hole without any mass ejection.  The
only evidence of this collapse will be the weak neutrino and
gravitational wave signal (see below).  The difference in fate of
these two collapsing stars highlights the sensitivity of the collapse
to the angular momentum profile in the core (a factor of ten in
angular momentum can make the difference between a fizzle and a
10$^{52}\,{\rm erg}$ explosion).

Our 3-dimensional simulations paint a very different picture than what
was obtained from 2-dimensional simulations.  Both the realistic
stellar structure and the fact that we are modelling the collapse in
three dimensions lead to the development of spiral-wave instabilities;
fluid viscosity ultimately transports energy outward through the star
and drives a hypernova-like explosion.  The ``disk'' formed in these
calculations does not fit the structure of any analytic studies
performed to date on GRB disks.  This means that our understanding of
how collapsars produce GRB jets as well as the nature of the
nucleosynthesis, neutrino emission and gravitational wave emission
from these stars may be substantially different than previous
predictions.  These changes will depend sensitively on the angular
momentum, and much more work must be done to fully understand
collapsars.

\begin{figure}
\plotone{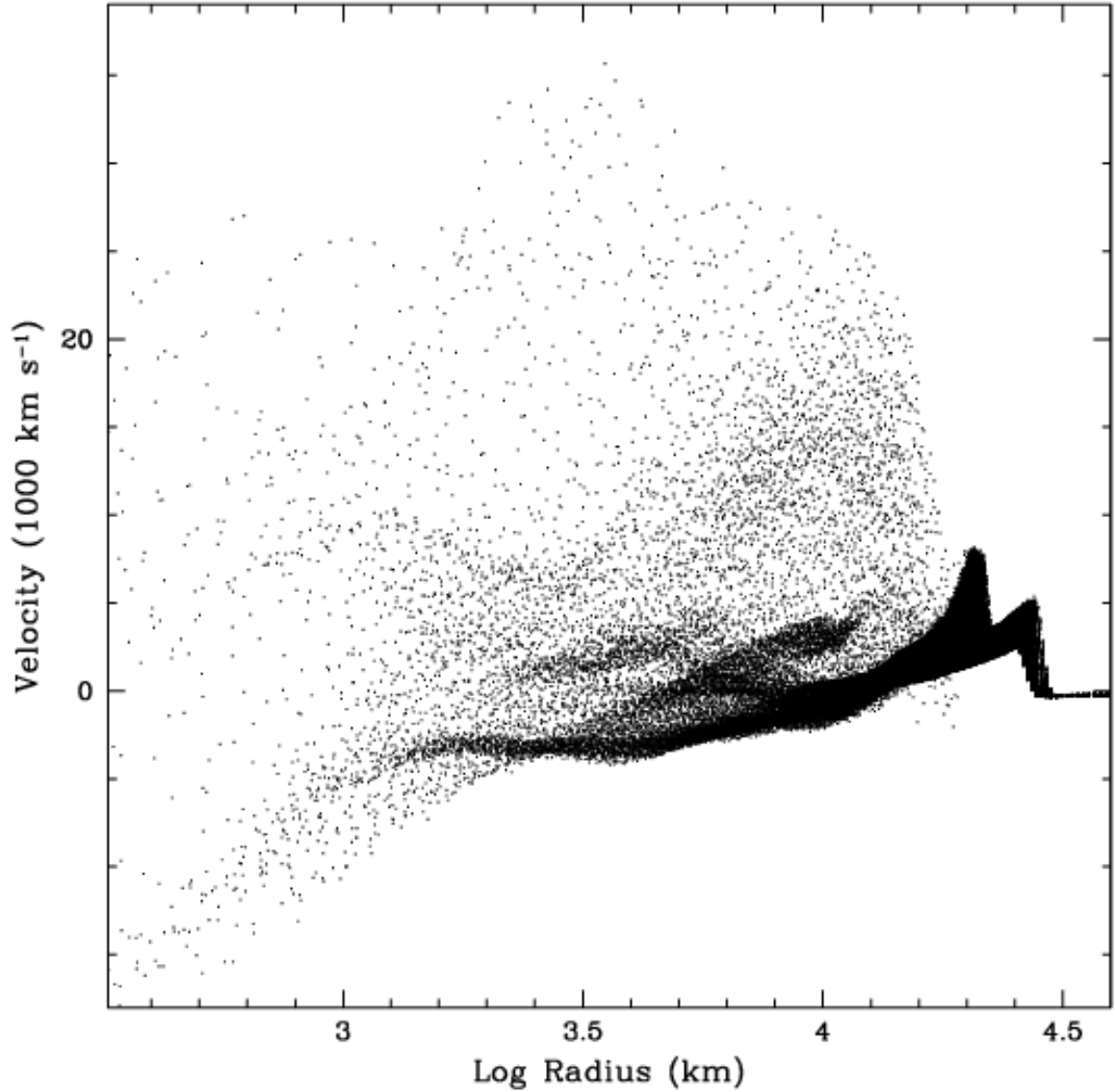}
\caption{Velocity of a wedge of material (material with $|y|/r <
  0.08$) as a function of radius 3.3\,s after collapse.  Two primary
  shocks exist, one near $5,000{\rm \, km s^{-1}}$ and the other near
  $8,000{\rm \, km s^{-1}}$.  But some material with velocities above
  $20,000{\rm \, km s^{-1}}$ also exists.  These high velocities will
  cause much of the star to go through explosive nucleosynthesis,
  probably producing considerable amounts of $^{56}$Ni.}
\label{fig:vel}
\end{figure}

\begin{figure}
\plotone{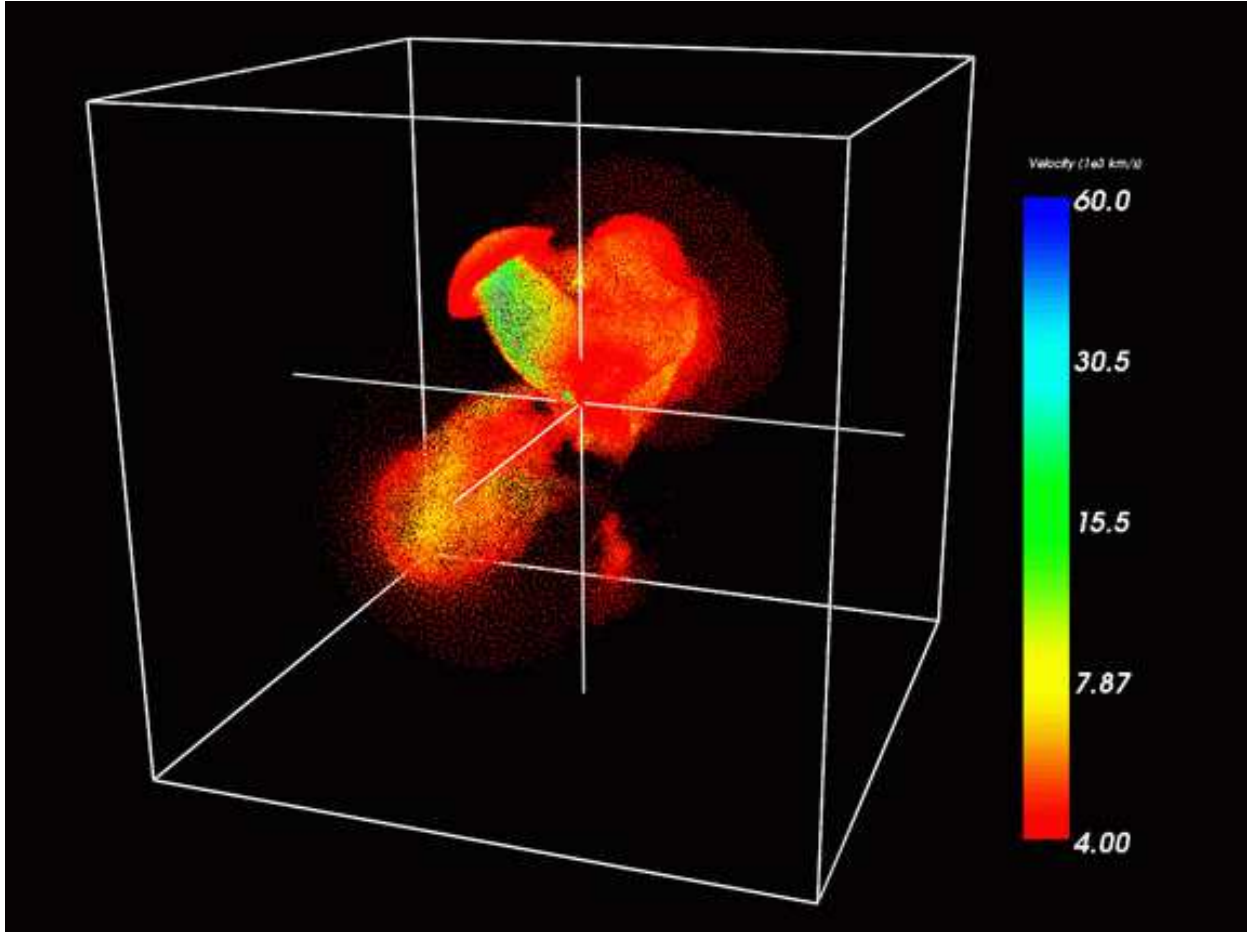}
\caption{SPH particles in the rapidly-rotating collapsar simulation,
  selected and colored by radial velocity, 7.8 seconds after collapse.
  Only particles with outward radial velocities greater than
  4000~km~s$^{-1}$ are shown in the image; areas containing no visible
  particles represent regions where the gas is moving outward
  relatively slowly, or actually falling inward toward the central
  black hole.  The half-side-length of the bounding box is 5.8e3~km.}
\label{fig:velasym}
\end{figure}

\begin{figure}
\plotone{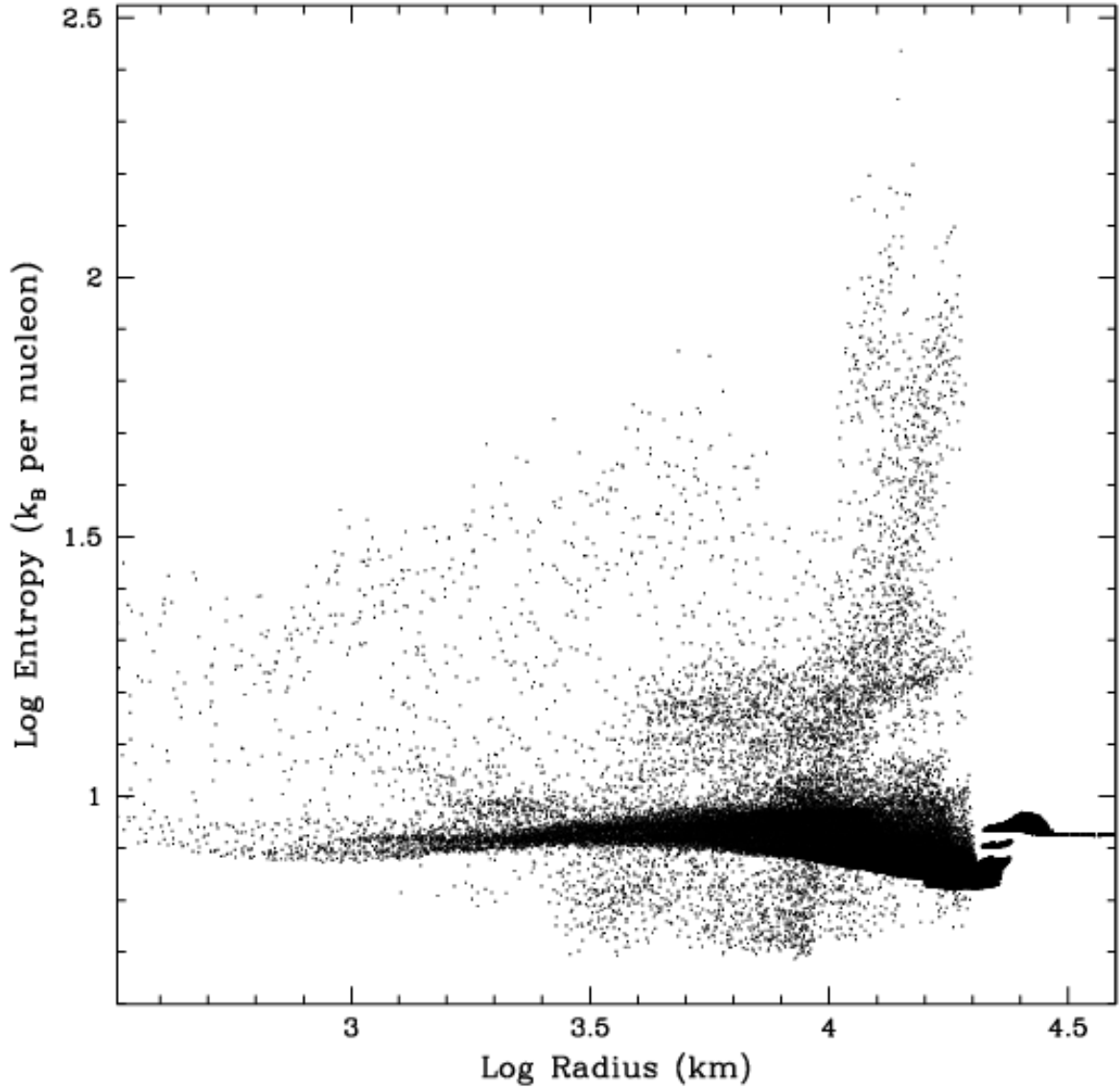}
\caption{Entropy of a wedge of material (material with $|y|/r < 0.08$)
  versus radius of the ejecta 3.3\,s after collapse.  Although much of
  the matter has normal/low entropies ($\sim 8-10$\,k$_{\rm B}$ per
  nucleon), some of the matter achieves extremely high entropies (in
  excess of $100$\,k$_{\rm B}$ per nucleon).  Such high entropies will
  alter the nucleosynthetic yields of this matter.}
\label{fig:entr}
\end{figure}

3.3\,s into the simulation, the ejecta from this viscosity-driven
explosion is moving at velocities in excess of $5,000-10,000 {\rm \,
  km \, s^{-1}}$ (Fig.~\ref{fig:vel}) with some ejecta moving faster
than $30,000 {\rm \, km \, s^{-1}}$.  At such high velocities, the
shock is strong enough to drive explosive nucleosynthesis well into
the star \citep[see, for example,][]{Mae03,Nom05,Fry06}.  The spiral
wave instability is imprinted on the motion of the ejecta and it is
far from symmetric (Fig.~\ref{fig:velasym}).  The entropy of this
shocked material is plotted in figure~\ref{fig:entr}.  Some of the
material reaches high entropies characteristic of hypernova explosions
(a small amount of matter reaches entropies in excess of 100\,k$_{\rm
  B}$ per nucleon) and these entropies will alter the signature of the
nucleosynthetic yield \citep[see][for details]{Mae03,Fry06}.  It is
likely that this explosive nucleosynthesis will dominate over disk
nucleosynthesis \citep*[e.g.][]{Sur06} for our simulation and
collapsar models like it.  We defer investigation of the exact yields
of this explosion to a later paper.

\begin{figure}
\plotone{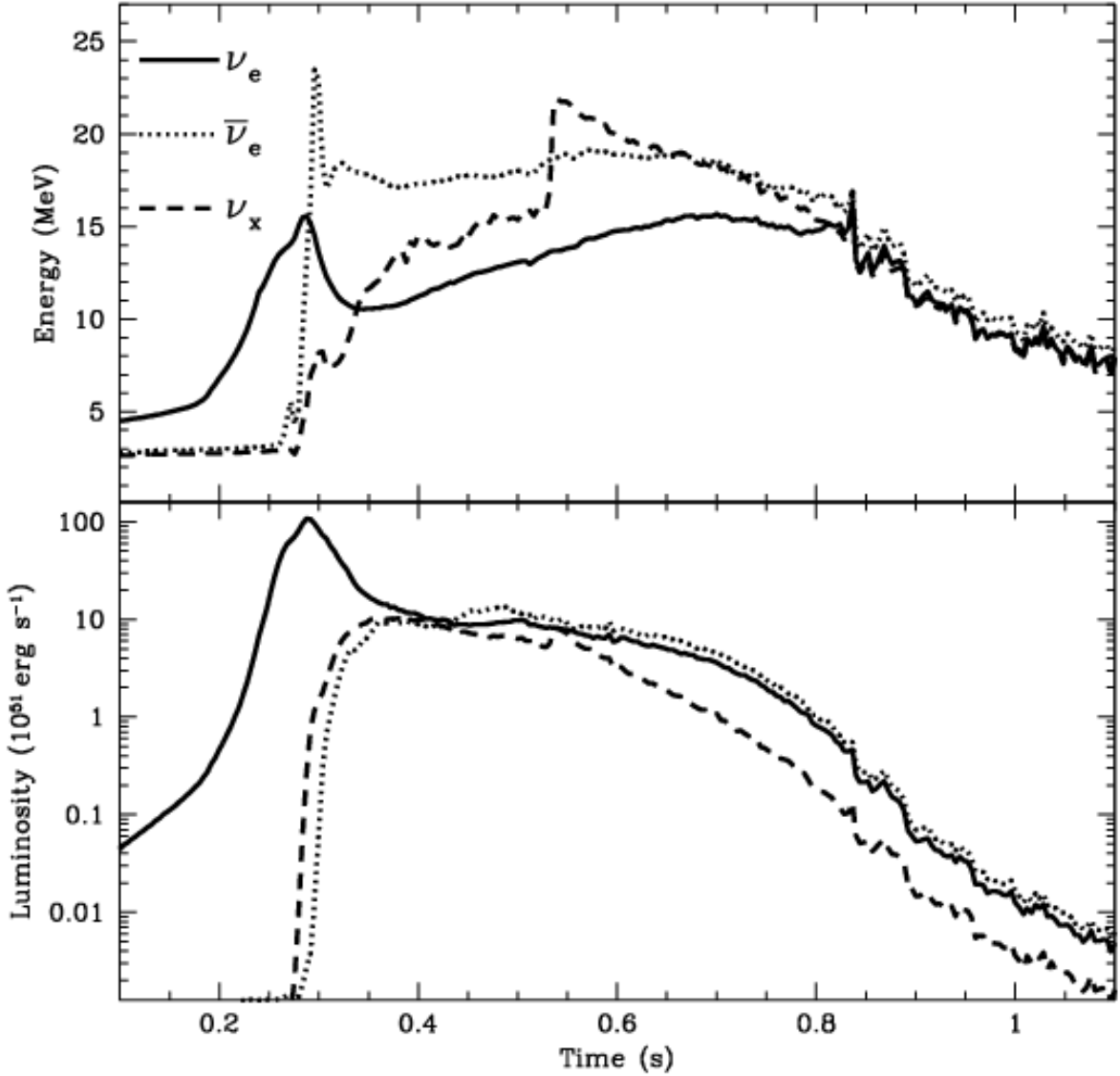}
\caption{The neutrino energy (top) and neutrino luminosity (bottom) as 
  a function of time in our simulation for the three neutrino flavors
  followed in our simulation: electron neutrino ($\nu_e$ - solid),
  electron anti-neutrino ($\bar{\nu}_e$ - dotted), and $\mu$ and
  $\tau$ neutrinos and anti-neutrinos ($\nu_x$ - dashed).  Note that
  the electron neutrino and anti-neutrino luminosities and energies
  are nearly equal at late times, suggesting that the neutrino flux
  will strive to set the electron fraction to 0.5 at these times.  The
  electron fraction is critical in determining the yield from matter
  ejected in this collapse.}
\label{fig:neut}
\end{figure}

Some of the ejecta will have low electron fractions.  The electron
neutrino and anti-neutrino fluxes very quickly evolve to be nearly
identical, but until the explosion cuts off the accretion, the
anti-neutrino energies are roughly 20\% higher (Fig.~\ref{fig:neut})
and hence will be absorbed preferentially over the electron neutrinos.
If the neutrino fluxes could reset the electron fraction of the
ejecta, the higher electron anti-neutrino absorption rate would
produce an electron fraction below 0.5.  However, bear in mind that
(1) the neutrino energies may be different with better transport
schemes, and (2) this enhanced emission occurs primarily when the
compact remnant is still likely to be a neutron star.  If our
simulation included the emission from electron capture in this neutron
star, the electron neutrino flux would be higher.  Although there are
some differences between the neutrino light-curves in this calculation
and that of a typical core-collapse supernova, most of these
differences occur because we have not included the neutron star prior
to collapse.

\begin{figure}
\plotone{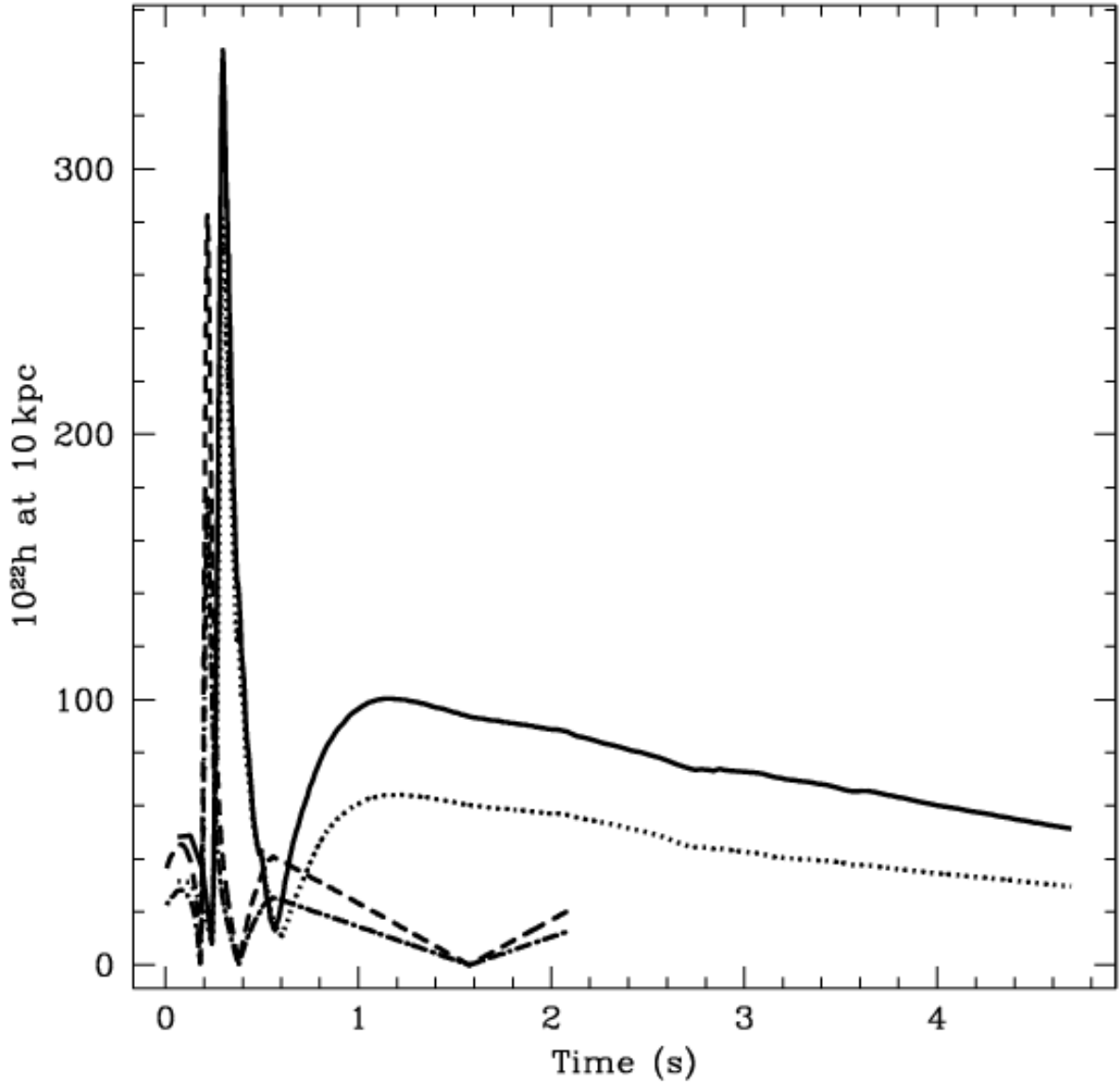}
\caption{Angle-averaged wave amplitude of the gravitational wave
  emission arising from mass motions as a function of time for the
  rapidly-rotating (solid: ${\langle h_\times^2 \rangle}^{1/2}$,
  dotted: ${\langle h_+^2\rangle}^{1/2}$) and slowly-rotating (dashed:
  ${\langle h_\times^2}\rangle^{1/2}$, dot-dashed: 
  ${\langle h_+^2\rangle}^{1/2}$) simulations.}
\label{fig:gwmm}
\end{figure}

One of the new discoveries in this study was the development of
instabilities in the disk.  The hot, self-gravitating torus of
material around the central black hole is susceptible to the formation
of spiral waves, and these waves facilitate enhanced outward transport
of angular momentum through the star.  The instabilities that develop
in the disk lead to mass motion and neutrino emission asymmetries that
can produce gravitational wave emission.  The frequency of these
oscillations is a function of the motion of the spiral wave
instability.  Although at the inner edge of the disk, oscillation
periods can be less than 1\,ms, in practice, the period has
frequencies in the 100-1000 Hz regime.  Using the formulism described
in \citet*{Fry04} for SPH data, we have calculated the amplitude of
the gravitational wave emission from both mass-motion and neutrino
asymmetries.  Figure~\ref{fig:gwmm} shows the signal from mass-motion
for our rapidly- and slowly-rotating models.  The maximum amplitude of
the signal, at the time of initial collapse, can be nearly 15 times
larger for this collapsar model than the strongest signal arising from
the rotating models in \citet{Fry04}.

\begin{figure}
\plotone{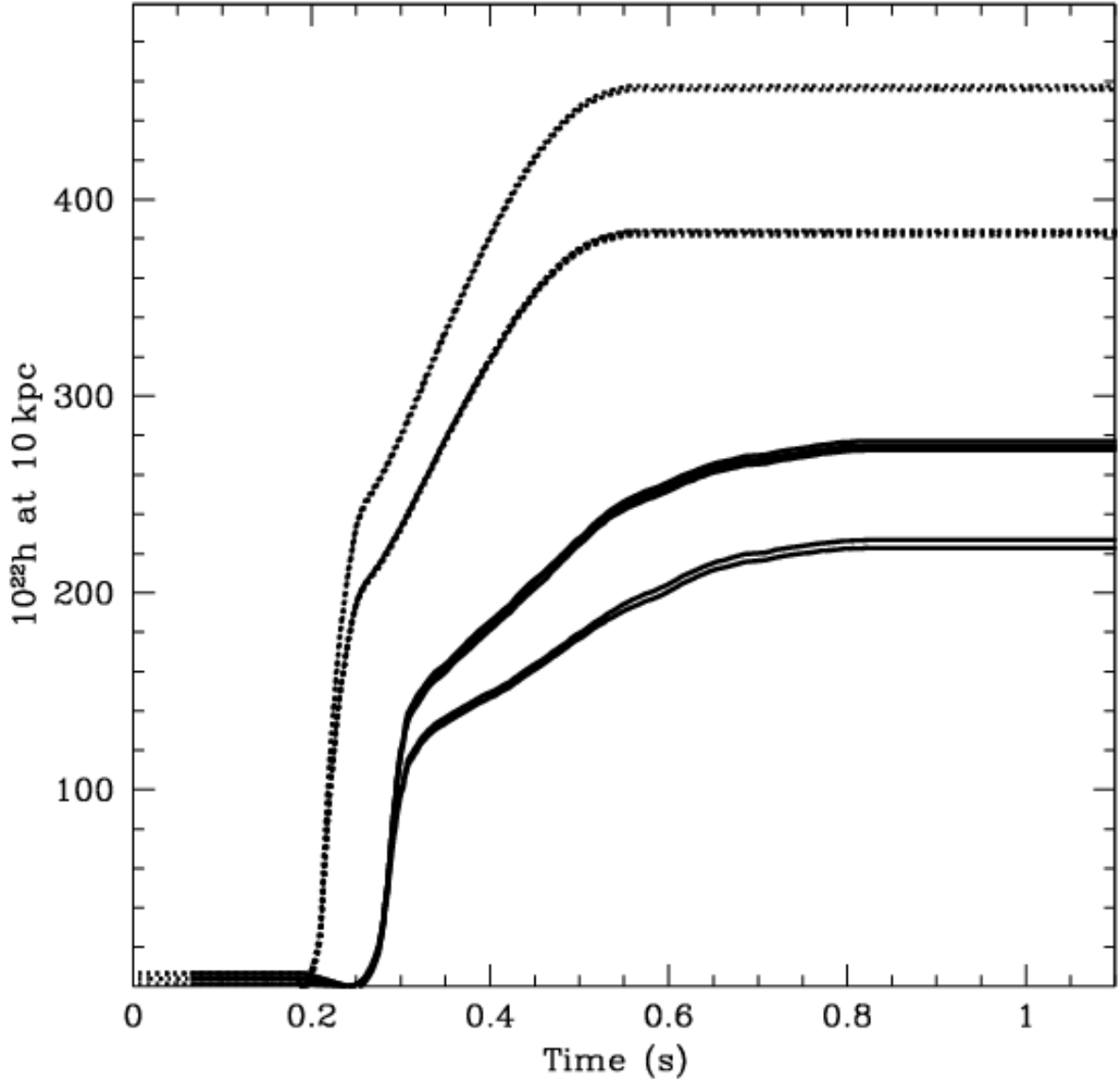}
\caption{The amplitude of the gravitational wave emission arising 
from neutrino asymmetries as a function of time for the rapidly-rotating 
(solid) and slowly-rotating (dotted) simulations.  The multiple lines denote 
different viewing angles and bracket the range of possible signals.}
\label{fig:gwnu}
\end{figure}

Comparing figure~\ref{fig:gwnu} to the work of \citet{Fry04}, we find
that the gravitational wave signal from asymmetric neutrino emission
is not significantly higher than what we would expect from supernovae.
In the case of our rapidly-spinning star, this is because the mass
outflows limit the amount of material that piles up around the black
hole, causing the neutrino luminosity to drop (Fig.~\ref{fig:neut}).
In the case of the slowly-spinning star, the neutrino luminosity is
never very asymmetric.  Even so, the neutrino gravitational wave
signal is at least as strong as that for supernovae (and could be a
factor of 5 higher in some cases).

With such results, we find it unlikely that LIGO or advanced LIGO can
detect the gravitational wave signal from collapsars out to 1\,Mpc,
making this class of GRBs (possibly all long-duration GRBs), with
their low occurrence rate, a poor source for gravitational wave
detectors.  However, we are not considering all the potential
gravitational wave emission mechanisms in the collapsar model; for
example, we do not consider black hole ringing, although Fryer et al.
(2002) argued this would be small.  We also do not model the actual
black hole formation, and this will change the signal considerably.

{\bf Acknowledgments} This work was carried out under the auspices of
the National Nuclear Security Administration of the U.S.\ Department
of Energy at Los Alamos National Laboratory under Contract No.\ 
DE-AC52-06NA25396.  This work was funded in part by a DOE SciDAC grant
DE-FC02-01ER41176, by a NASA grant SWIF03-0047, and by the National
Science Foundation under Grant No.\ PHY99-07949.

{}

\clearpage
\newpage

\clearpage

\clearpage

\clearpage

\clearpage

\clearpage

\clearpage

\clearpage

\begin{figure}

\plottwo{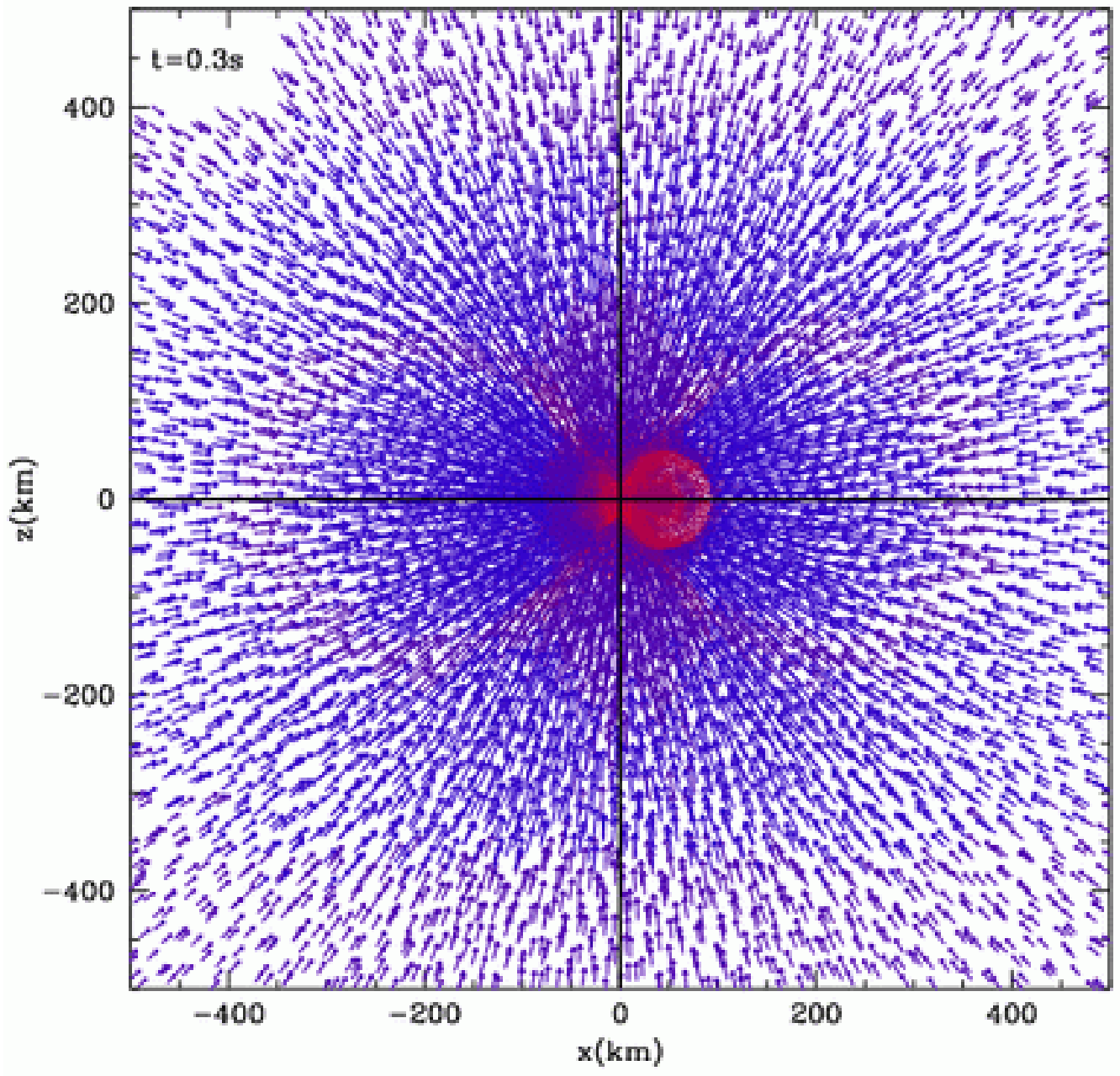}{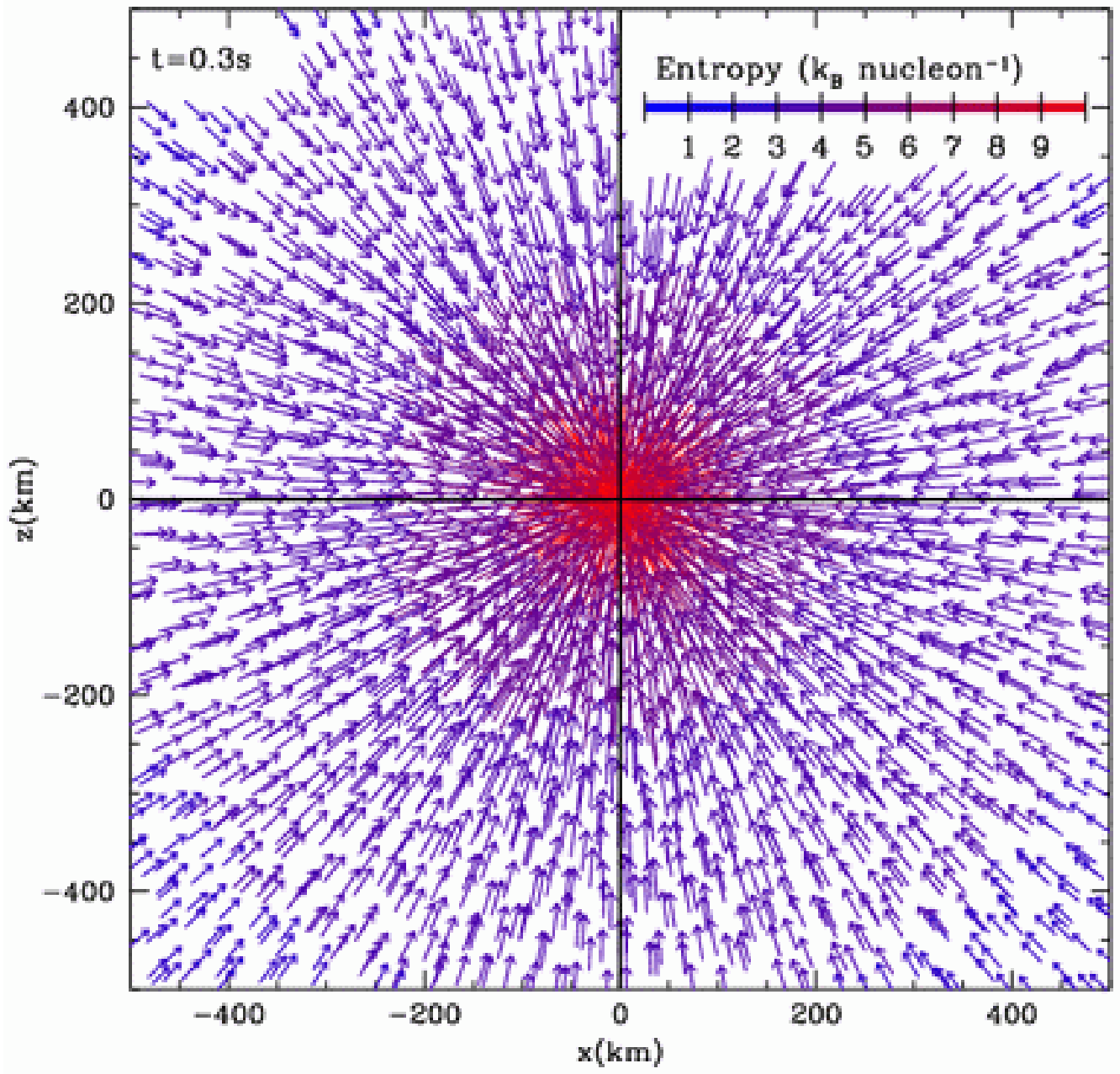}

\plottwo{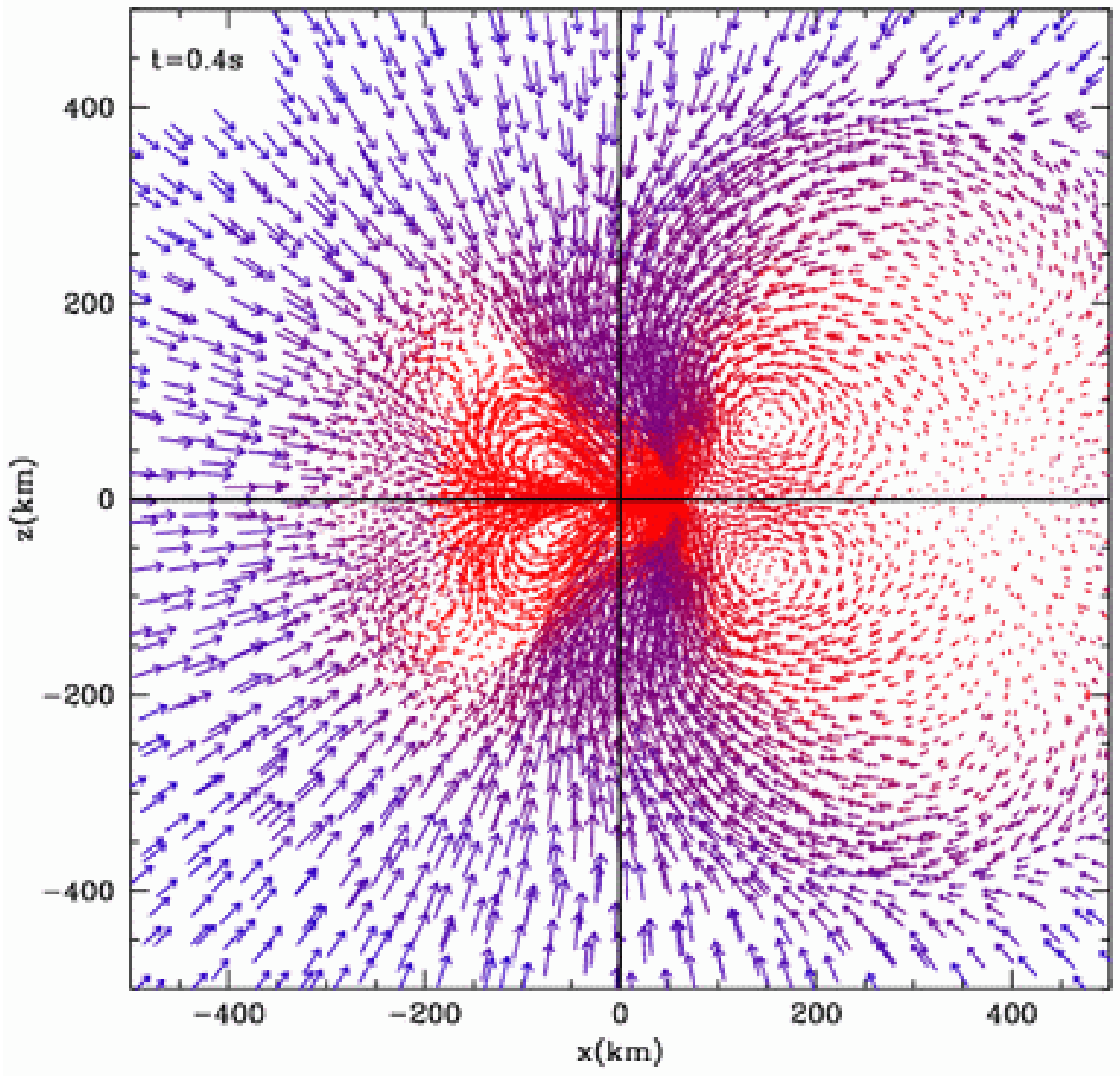}{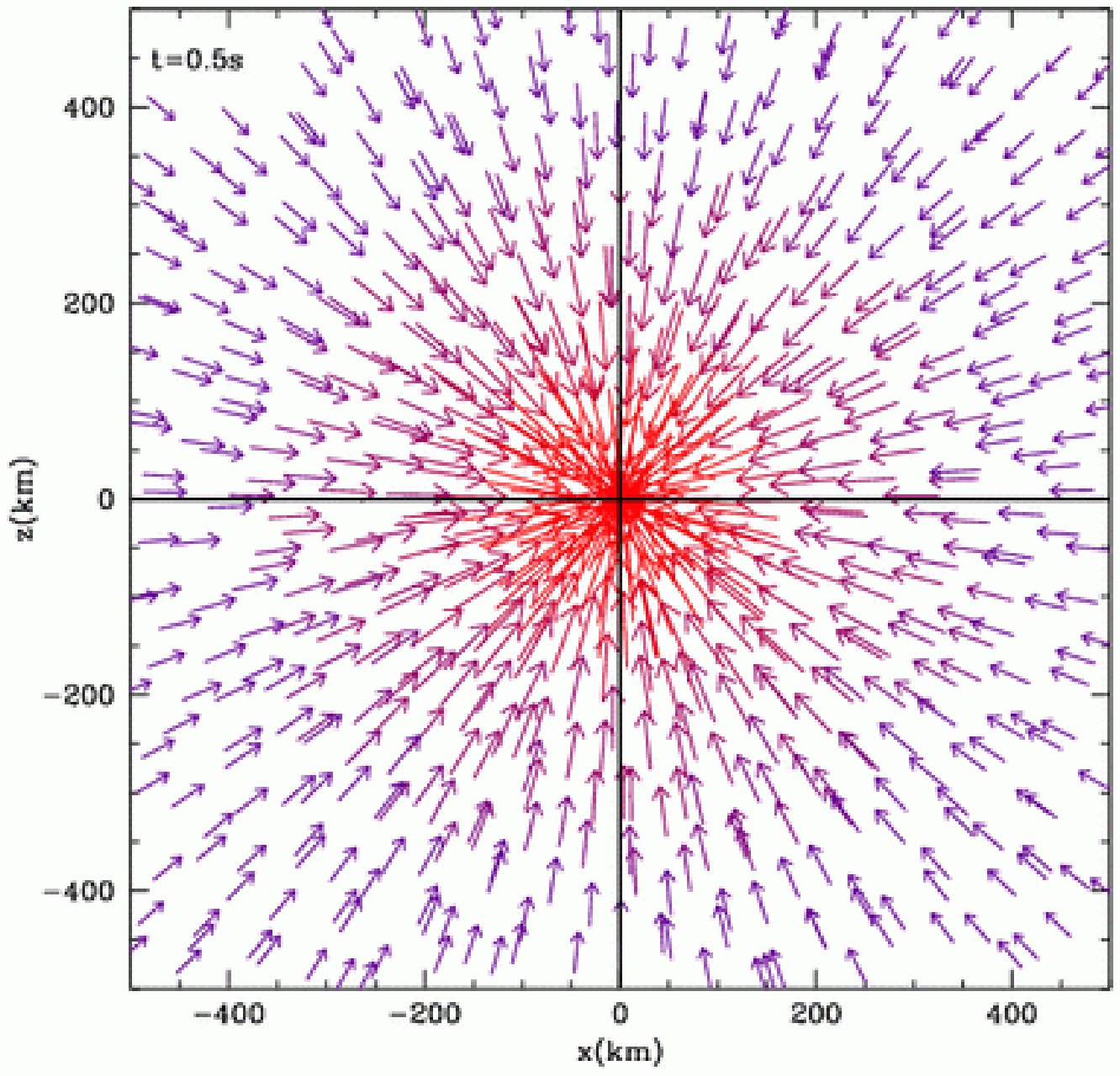}

\end{figure}
\clearpage

\clearpage

\clearpage

\clearpage

\clearpage

\clearpage

\clearpage

\clearpage

\clearpage

\end{document}